\documentstyle[prl,aps,floats,psfig,epsf]{revtex}
\catcode`@=11

\draft

\def\lsim{\mathrel{\raise.3ex\hbox{$<$\kern-.75em\lower1ex\hbox{$\sim$}}}}
\def\gsim{\mathrel{\raise.3ex\hbox{$>$\kern-.75em\lower1ex\hbox{$\sim$}}}}

\begin{document}

\twocolumn[\hsize\textwidth\columnwidth\hsize\csname
@twocolumnfalse\endcsname

\hfill\vbox{
\hbox{BUHEP-03-05}
\hbox{MADPH-03-1321}
\hbox{NSF-KITP-03-16}
\hbox{hep-ph/0302150}
\hbox{}}

\title{WMAP and Inflation}
\author{$^{1,3}$V. Barger, $^1$Hye-Sung Lee and $^{2,3}$Danny Marfatia}
\vskip 0.3in
\address{$^1$Department of Physics, University of Wisconsin, Madison, WI 53706}
\vskip 0.1in
\address{$^2$Department of Physics, Boston University, Boston, MA 02215}
\address{$^3$Kavli Institute for Theoretical Physics, University of
California, Santa Barbara, CA 93106}
\maketitle

\begin{abstract}

We assay how inflationary models whose properties are dominated by the 
dynamics of a single scalar field are constrained by 
cosmic microwave background (CMB) data from the Wilkinson 
Microwave Anisotropy 
Probe (WMAP). We classify inflationary models in a plane defined by the 
horizon-flow parameters. 
Our approach differs from that of the WMAP collaboration in that
we analyze only WMAP data and take the spectral shapes from 
slow-roll inflation rather than power-law parameterizations of the spectra. 
The only other information we use is the measurement 
of $h$ from the Hubble Space Telescope (HST) Key Project.
We find that the spectral
index of primordial density perturbations lies in the $1\sigma$ range
$0.94\leq n_s\leq 1.04$ with no evidence of running.
The ratio of the amplitudes of tensor and scalar perturbations is smaller
than $0.61$ and the inflationary scale is below
2.8$\times 10^{16}$ GeV, both at the 2$\sigma$ C.~L. 
No class of inflation or ekpyrotic/cyclic model is excluded. The
 $\lambda \phi^4$ potential is excluded at $3\sigma$ 
only if the number of e-folds is assumed
to be less than 45.

\end{abstract}
]

That there are multiple peaks in the CMB has recently been reinforced 
 by data from the WMAP satellite~\cite{map,maptt,mapte}. 
This establishes that the curvature fluctuations which seed 
structure formation were generated at superhorizon scales. The inflationary
paradigm~\cite{inflation}, 
which hinges on this very fact~\cite{superhorizon}, is therefore vindicated.
Other generic predictions of inflation, including approximate 
scale-invariance of the power spectra, flatness of the Universe and 
adiabaticity and gaussianity of the density perturbations are also
fully consistent with WMAP data~\cite{map2}.

Given this supporting evidence for inflation, we adopt the full set of
predictions of slow-roll inflation and obtain constraints on inflationary 
models imposed by WMAP's data. Studies of this type have been carried out
in the past with less precise data~\cite{leach} 
and with the simple parameterization 
of the power spectra as power laws~\cite{kinney}. 
It has been emphasized that to extract
precise information from data of WMAP's quality, high-accuracy predictions 
of the power spectra resulting from slow-roll inflation should be 
used~\cite{b}. There is a wealth of cosmological information in 
the CMB~\cite{hu}, 
and our approach is to employ precise theoretical 
expectations in its extraction.

Since the WMAP collaboration has considered what implications their data
have for inflation~\cite{map1}, we describe at the outset the differing 
elements between our analysis and theirs. We elaborate on these differences
later.
WMAP include CBI~\cite{cbi}, ACBAR~\cite{acbar}, 
2dFGRS~\cite{2df} and Lyman-$\alpha$ power spectrum~\cite{lyman}
 data in addition to their own data. We
restrict ourselves to WMAP data with a top-hat prior on the Hubble constant 
$h$ ($H_0=100 h$ km/s/Mpc), from 
the HST~\cite{HST}. 
While we use the actual theoretical 
predictions for the primordial power spectra from single-field slow-roll
inflation, WMAP parameterize the spectra
with power-laws and a running spectral index. As a result, the spectral
shapes we use are different from those of WMAP and we  directly fit to
slow-roll parameters, while WMAP fit to derivative quantities. There
are different virtues of the two approaches, and a comparison of the results 
obtained from them serve as a check of the robustness of the conclusions
reached. 

\vskip 0.1in
\noindent
{\bf Primordial spectra}:

The primordial scalar and tensor power spectra to ${\cal O}(\ln^2 k)$ 
are{\footnote{$\chi$ is
the intrinsic curvature perturbation and $h_{ij}$ is the transverse traceless
part of the metric tensor.}}~\cite{b,c}
\begin{eqnarray}
\label{eq:chi}
P_\chi&=&A_s\big(a_0+a_1\ln {k \over k_\star}+a_2\ln^2 
{k \over k_\star}\big)\,,\\
P_h&=&A_t \label{eq:h}
\big(b_0+b_1\ln {k \over k_\star}+b_2\ln^2 
{k \over k_\star}\big)\,,
\end{eqnarray}
where the pivot $k_\star$ typifies scales probed by the CMB. The constants
$a_i$ and $b_i$ are functions~\cite{b,a} of the horizon-flow parameters, 
$\epsilon_i$ of Ref.~\cite{horizon}, that are defined by
\begin{eqnarray}
\epsilon_{i+1}&=&{d \ln |\epsilon_i| \over d N}\,, \ \ \ i\geq 0 \\
 \epsilon_{0}&=& {H_I \over H}\,.
\end{eqnarray} 
Here, $N$ is the number of e-folds since some moment $t_I$ 
during inflation, 
when the Hubble parameter was $H_I$. 

Note that to ${\cal O}(\epsilon^2)$, the $b_i$ depend only on $\epsilon_1$
and $\epsilon_2$, while the $a_i$ depend on  $\epsilon_1$, $\epsilon_2$
and $\epsilon_3$.
We initially set $\epsilon_3=0$, but we then later demonstrate that the 
fit to the WMAP data is essentially not improved by including nonzero 
$\epsilon_3$.
The reason for not simply using the 
$\epsilon_3$-independent ${\cal O}(\epsilon)$ expressions is that
the ${\cal O}(\epsilon^2)$ expressions are more
accurate far from the pivot,
and for a wider range of $\epsilon_1$ and $\epsilon_2$~\cite{leach,b}.
It is uncertain that even high-precision data from the Planck
satellite~\cite{planck} can constrain $\epsilon_3$ to be small. 
Including $\epsilon_3$ in our analysis 
would simply enlarge the allowed 
parameter space to include models which are not
inflationary in the sense that
the horizon-flow parameters are not small.

The primary advantage of the horizon-flow parameters is that accurate 
predictions of 
the shapes and normalizations of the power spectra can be made
 independent of parameters describing cosmic evolution. 
The horizon-flow parameters $\epsilon_1$ and $\epsilon_2$
 are related to the usual slow-roll
parameters~\cite{lidsey}
\begin{eqnarray}
\epsilon&\equiv&{M_{Pl}^2 \over 16 \pi} \big({V'\over V}\big)^2\,,\\
\eta &\equiv& {M_{Pl}^2 \over 8 \pi} \bigg[{V''\over V}-{1\over 2}\big({V'\over V}\big)^2\bigg] \,,   
\end{eqnarray} 
via the first order relations~\cite{b}
\begin{equation}
  \epsilon_1\simeq \epsilon\,,\ \ \ \ \ \ \ \epsilon_2\simeq 2(\epsilon-\eta)\,.
\end{equation}
The normalizations of the spectra are given by
\begin{equation}
  \label{eq:amplitude}
  A_s={H_I^2\over \pi \epsilon_1 M_{Pl}^2}\,,\ \ \ \ \ \ \ 
  A_t={16 H_I^2 \over \pi M_{Pl}^2}\,,
\end{equation}
and the ratio of the amplitudes of the spectra at the pivot $k=k_\star$, is
\begin{eqnarray}
  \label{eq:R}
 R&\equiv& {A_t b_0 \over A_s a_0} =16 \epsilon_1\big[1+C\epsilon_2 \nonumber 
\\ &&
+\big(C-{\pi^2 \over 2}+5\big)\epsilon_1 \epsilon_2+
\big({C^2 \over 2}-{\pi^2 \over 8}+1\big)\epsilon_2^2\big]\,,
\end{eqnarray}
where $C \equiv \gamma_E+\ln 2-2 \approx -0.7296$. Note that $a_0$ and $b_0$
are ${\cal O}(1)$ and  $|a_0-b_0|$ is ${\cal O}(\epsilon_2)$.
The spectral indices and their running can be expressed in
terms of $\epsilon_1$, $\epsilon_2$ and $\epsilon_3$:
\begin{eqnarray}
n_s& = &1-2\epsilon_1-\epsilon_2-2\epsilon_1^2-(2C+3)\epsilon_1 \epsilon_2
-C\epsilon_2 \epsilon_3\,,\\
n_t &=& -2\epsilon_1-2\epsilon_1^2-2(C+1)\epsilon_1 \epsilon_2\,,\\
\label{as}
  \alpha_s& \equiv & {d n_s \over d \ln k}=-2\epsilon_1 \epsilon_2-
\epsilon_2 \epsilon_3\,,\\ 
\label{at}
\alpha_t&\equiv &{d n_t \over d \ln k}=-2\epsilon_1 \epsilon_2\,.
\end{eqnarray}
We set $\epsilon_i=0$, $i\geq 3$.
$R$, $n_s$ are $n_t$ are interdependent and the following
consistency condition on 
single-field slow-roll inflation applies:
\begin{equation}
\label{consistency}
R \simeq -8 n_t  
\end{equation}
By our choice of formalism, we implicitly assume this condition 
to be satisfied.

Note that the six inflationary parameters, $A_s$, $A_t$, $n_s$, $n_t$, 
$\alpha_s$ and $\alpha_t$ are determined by just three parameters, $A_s$,
$\epsilon_1$ and $\epsilon_2$ in our analysis. In contrast, the WMAP 
collaboration parameterize the power-spectra with~\cite{map1}
\begin{equation}
  \label{eq:pl+r}
  P_\chi=A_s \big({k \over k_\star}\big)^{n_s-1+{1\over 2}\alpha_s \ln{k \over k_\star}}\,, \ \ \ \
P_h=A_t \big({k \over k_\star}\big)^{n_t}\,.
\end{equation}
They eliminate $n_t$ as a free parameter by using 
the consistency condition Eq.~(\ref{consistency}). Thus, they
 have four free parameters, $A_s$, $A_t$, $n_s$ and $\alpha_s$ 
and set $\alpha_t=0$. 

A convenient classification of models based on the separate 
regions in the $R-n_s$ 
plane they populate, or equivalently relationships between the slow-roll 
parameters, was introduced in Ref.~\cite{classification}. 
This classification becomes 
particularly simple in the $\epsilon_2-\epsilon_1$ plane, as discussed in the
following section
and shown in Fig.~\ref{fig:1}.

\vskip 0.1in
\noindent
{\bf Inflation models}:

In what follows we use the common 
jargon, ``red-tilt'' for $n_s<1$ and ``blue-tilt'' for $n_s>1$.

\noindent
a) 
Canonical potentials of {\it large-field} models are the monomial potential, 
\begin{equation}
  \label{eq:large}
  V(\phi)= V_0 (\phi/\mu)^p\,, \ \ \ \ p\geq 2\,,
\end{equation}
and the exponential potential 
$V(\phi)= V_0 e^{\phi/\mu}$ of power-law inflation. They are typical 
of chaotic inflation~\cite{chaotic} and have $V''>0$. 
The value of the scalar field falls  ${\cal O}(M_{Pl})$
while the relevant perturbations are generated and thereby 
offer a glimpse into Planckian physics.
In terms of the horizon-flow parameters, large-field models satisfy 
\begin{equation}
  0\leq \epsilon_2<4\epsilon_1\,.
\end{equation}
They have large $R$ and predict a red tilt.

\noindent
b)
Generic {\it small-field} 
potentials are of the form 
\begin{equation}
  \label{eq:small}
  V(\phi)= V_0 [1-(\phi/\mu)^p]\,, \ \ \ \ p\geq 2\,,
\end{equation}
 and are therefore characterized by $V''<0$. The scalar field
rolls from an unstable equilibrium at the origin towards a non-zero vacuum 
expectation value. Models relying on spontaneous symmetry breaking yield
such potentials~\cite{sym}. For  small-field models
\begin{equation}
  \label{eq:sm}
  \epsilon_2> 4\epsilon_1\,. 
\end{equation}
The tensor fraction is small and the scalar spectrum is red-tilted. 

\noindent
c)
Potentials for {\it hybrid}
inflation~\cite{hybrid} are of the form 
\begin{equation}
  \label{eq:hybrid}
 V(\phi)= V_0 [1+(\phi/\mu)^p]\,, \ \ \ \ p\geq 2\,. 
\end{equation}
Hybrid inflation
models involve multiple scalar fields. One of the fields, $\phi$,
 is the slowly rolling 
inflaton which does not carry most of the energy density 
($\phi<\mu$). 
 Another field which has a fixed value during the slow-roll of $\phi$
provides $V_0$. When $\phi$ falls below a critical value, the other field
is destabilized and promptly ends inflation. 
As a result the value of $\phi$ at the
end of inflation is very model-dependent, which makes the number of e-folds
correspondingly uncertain. Hybrid models can be treated 
as single-field models because the only role of the second field is 
to end inflation, and the slow-roll dynamics is dominated by a single field. 
These potentials arise in
supersymmetric and supergravity models of inflation. 
Hybrid models have
\begin{equation}
  \label{eq:hy}
  \epsilon_2<0\,.
\end{equation}
There is no robust prediction for $R$ as can be seen from the above
$\epsilon_1$-independent inequality.
However, a unique prediction 
of these models is that the spectrum is blue-tilted 
if $|\epsilon_2|>2\epsilon_1$.

The line 
$\epsilon_2=0$ ($R=8(1-n_s)$) implies $\epsilon=\eta$ which occurs for the 
exponential potential. Thus, power-law inflation
marks the boundary between large-field and hybrid models. 

\noindent
c)
{\it Linear} potentials, 
\begin{equation}
  \label{eq:linear}
  V(\phi)= V_0 (\phi/\mu)\,,\ \ \ \  V(\phi)= V_0[1-(\phi/\mu)]\,,
\end{equation}
define the boundary between large and small field models and lie at
\begin{equation}
  \label{eq:lin}
  \epsilon_2=4\epsilon_1\, \ \ \ {\rm{or}} \ \ \ 3R=8(1-n_s)\,. 
\end{equation}
Since $\epsilon_1$ is a constant for such potentials,
inflation ends only with the help of an auxiliary field or some other physics.

To avoid overstating the comprehensiveness of this classification of 
potentials, we list a few potentials of different forms~\cite{review}, 
which however
do fit into the large-field, small-field, hybrid or linear categories 
according to the relationships between the horizon-flow parameters. 
$V(\phi)=V_0 [1 \pm \ln(\phi/\mu)]$, $V_0 [1-e^{-\phi/\mu}]$ and 
 $V_0 [1\pm (\phi/\mu)^{-p}]$
are hybrid in the sense that an auxiliary field
is needed to end inflation, but lead to 
$\epsilon_2>4\epsilon_1$ and therefore lie in the small-field region of the
parameter space. Similarly, power-law inflation does not end without a hybrid
mechanism, but lies in the large-field region. Finally, let us note that
the simplest models of the ekpyrotic/cyclic~\cite{ekpyrotic} variety are 
small-field~\cite{cyclic}; however, the prediction for $n_s$ in these models
is controversial~\cite{controversy}.

\vskip 0.1in
\noindent
{\bf Analysis}:

We compute the TT and TE power spectra $\delta T_l^2=l(l+1)C_l/2 \pi$, 
using the Code for Anisotropies in the 
Microwave Background or CAMB~\cite{camb} (which is a parallelized
version of CMBFAST~\cite{cmbfast}) and a supporting module that calculates the inflationary
predictions for the primordial scalar and tensor power spectra~\cite{code}.
We assume 
the Universe to be flat, in accord with the predictions of inflation, and that 
     the neutrino contribution to the matter budget is negligible. The dark
energy is assumed to be a cosmological constant $\Lambda$.
We calculate
the angular power  spectrum on a grid consisting of the parameters specifying 
the primordial spectra{\footnote{We call these 
\it{inflationary parameters.}}}, 
$\epsilon_1$, $\epsilon_2$ and $A_s$, and those 
specifying cosmic evolution{\footnote{We call 
these \it{cosmological parameters.}}}, 
the Hubble constant
$h$, the reionization optical depth $\tau$, the baryon
density $\omega_b= \Omega_B h^2$ and the total matter density 
$\omega_M= \Omega_M h^2$ (which is comprised of baryons 
and cold dark matter).
We choose $h$ rather than $\Omega_\Lambda$ because it is directly 
constrained by the HST~\cite{HST}. 
We do not include priors from supernova~\cite{sn}, gravitational
lensing~\cite{lensing} or large scale structure~\cite{lss} data
 or nucleosynthesis
constraints on $\omega_b$~\cite{BBN} although these
would somewhat sharpen the cosmological parameter determinations.

We employ the following top-hat grid:
\begin{itemize}
\item {$0.0001 \leq \epsilon_1 \leq 0.048$ in steps of size 0.004}
\item {$-0.18 \leq \epsilon_2 \leq 0.14$ in steps of size 0.02}
\item {$0.64 \leq h \leq 0.80$ in steps of size 0.02}
\item {$\tau=$0, 0.05, 0.1, 0.125, 0.15, 0.16, 0.17, 
0.18, 0.19, 0.2, 0.25, 0.3}
\item {$0.018 \leq \omega_b \leq 0.028$ in steps of size 0.001} 
\item {$0.06 \leq \omega_M \leq 0.22$ in steps of size 0.02}
\item {$A_s$ is a continuous parameter.}
\end{itemize}
The range of values chosen for $h$ correspond to the HST measurement,
$h=0.72\pm 0.08$~\cite{HST}. This serves to break the degeneracy between
$\Omega_M$ and $\Omega_\Lambda$ without the need for supernova 
data~\cite{break}. 
We place the pivot at $k_\star=0.007$ Mpc$^{-1}$. The
primordial spectra are evaluated to ${\cal O}(\ln^2 k)$ (see 
Eqs.~\ref{eq:chi}-\ref{eq:h}) with $\epsilon_3=0$. 

The WMAP data are in the form of 899 measurements of the TT power spectrum from
$l=2$ to $l=900$~\cite{maptt} and 449 data points of the TE power
spectrum~\cite{mapte}. 
We compute the  likelihood of each model of our grid using Version 1.1 of the
code provided by the collaboration~\cite{mapcode}. The WMAP code computes 
the full covariance matrix under the assumption that the off-diagonal terms are
subdominant. This approximation breaks down for unrealistically small
amplitudes. We include the off-diagonal terms 
only when the first peak occurs
 between $l=100$ and $l=400$ with a height above 
5000 $\mu K^2$~\cite{mapcode}.  
The restriction on the peak location may at first 
appear irrelevant, 
but for very large tensor amplitudes, the maximum height of the scalar 
spectrum shifts to very small $l$. 
When the height of the first peak is below 5000 $\mu K^2$
(which is many standard deviations away from the data), we only
use the diagonal terms of the covariance matrix to compute the likelihood.
 To obtain single-parameter 
constraints,
we plot the relative likelihood 
$e^{(\chi^2_{min}-\chi^2)/2}$, for each parameter after maximizing over all
the others. The $x$-$\sigma$ range is obtained for likelihoods above
$e^{-x^2/2}$. 
For 2-dimensional constraints, 
the $1\sigma$, $2\sigma$ and $3\sigma$ regions are defined by
$\Delta \chi^2=2.3$, 6.17 and 11.83, respectively, for two 
degrees of freedom.

\vskip 0.1in
\noindent
{\bf Results}:

Although our primary focus is to obtain constraints on the inflationary models,
we display constraints on the cosmological parameters
as a consistency check with the WMAP collaboration. This is
 pertinent because we have used a form
for the primordial spectra (Eqs.~\ref{eq:chi}-\ref{eq:h}) that is specific
to single-field slow-roll inflation rather than the standard 
power-law form. 

\begin{figure}[ht]
  \begin{center}
\mbox{\psfig{file=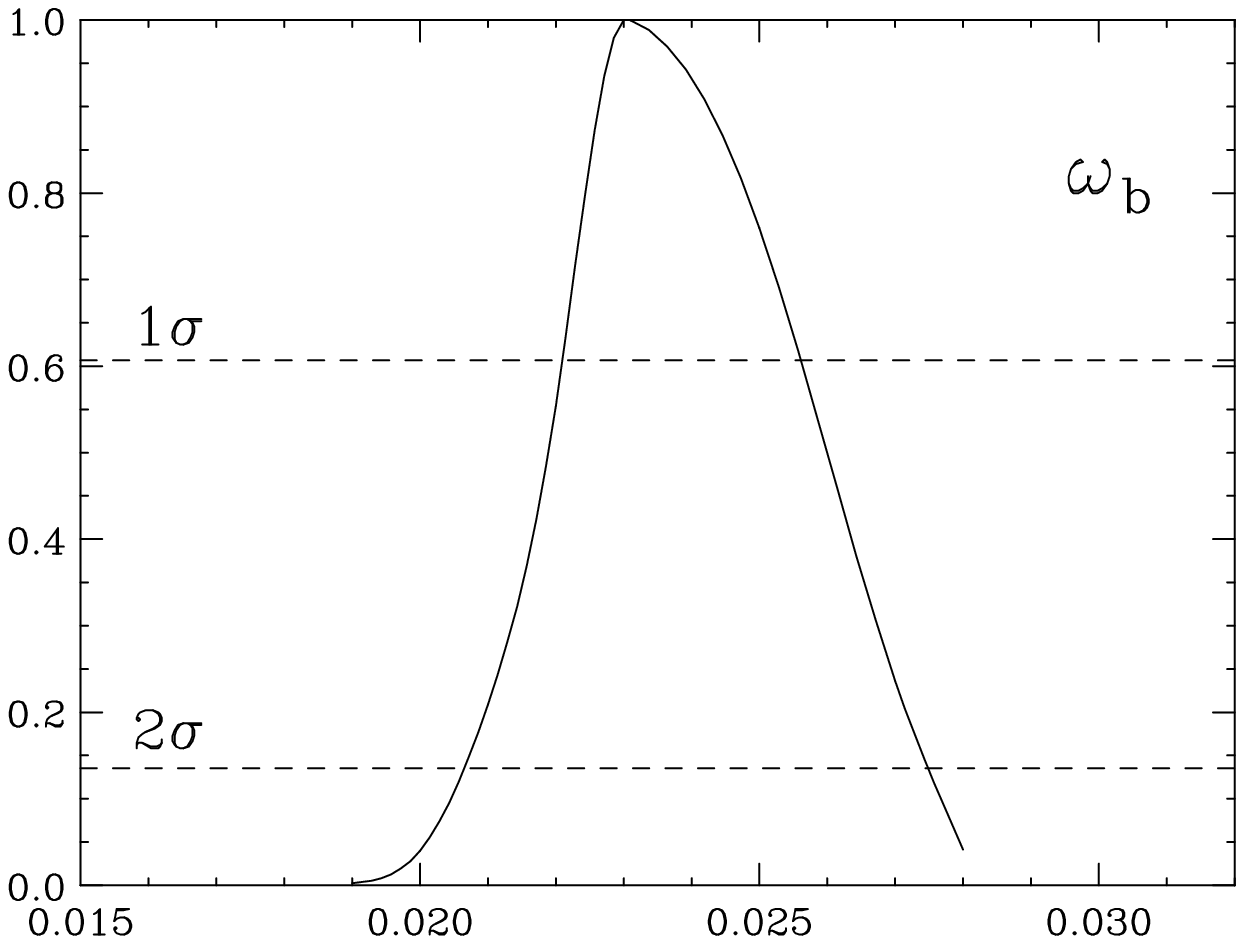,width=4.cm,height=3.5cm}
\psfig{file=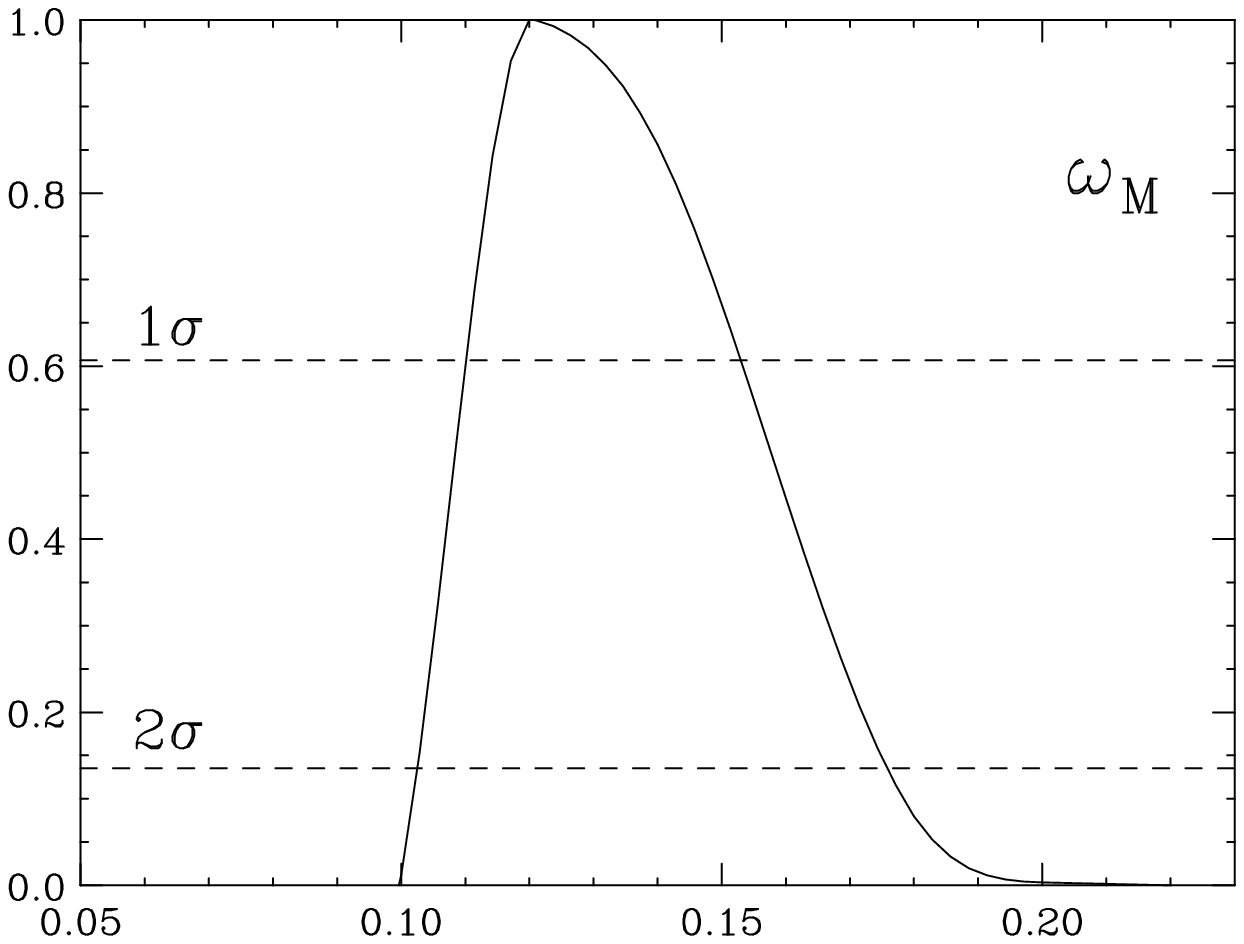,width=4.cm,height=3.5cm}}
\mbox{\psfig{file=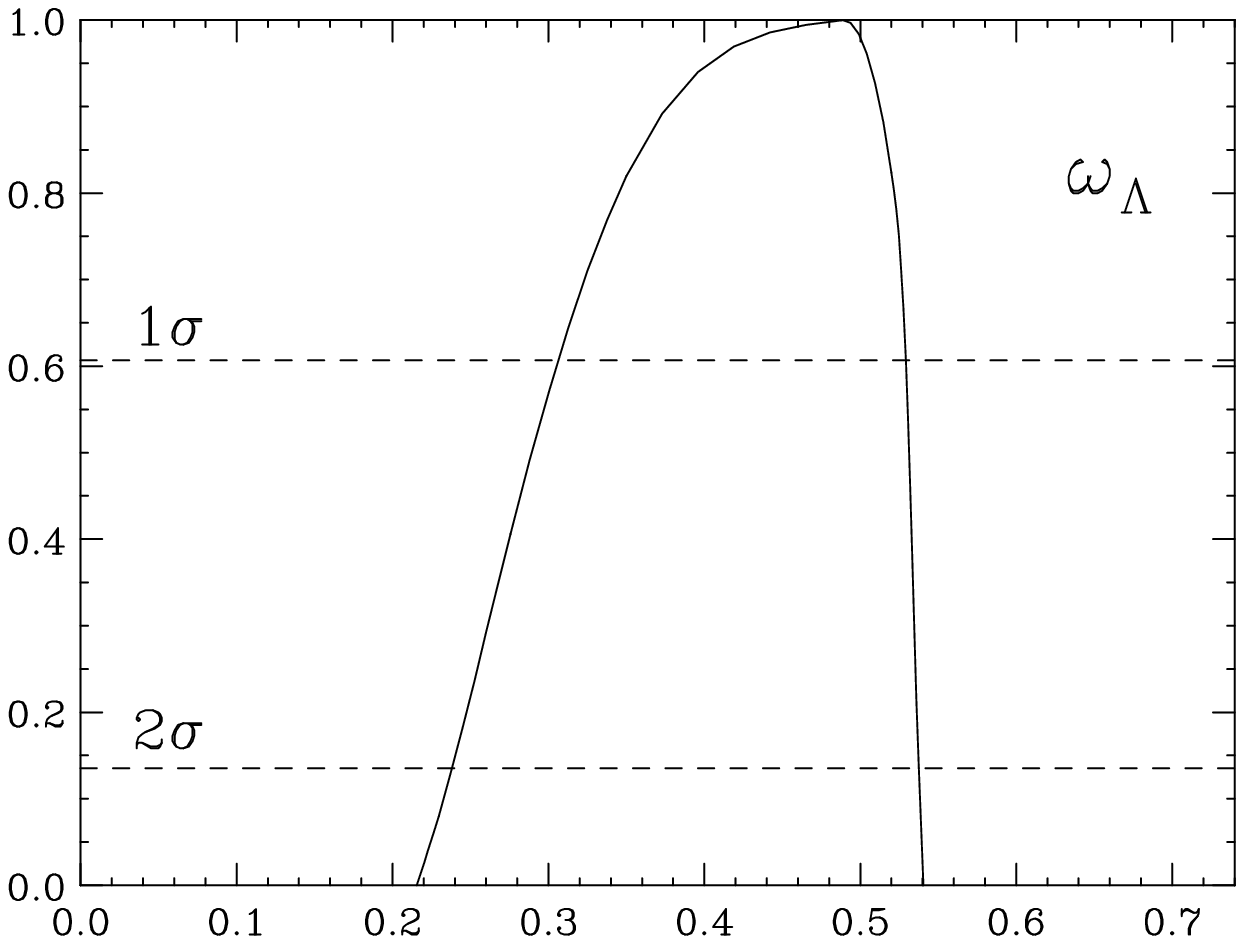,width=4.cm,height=3.5cm}
\psfig{file=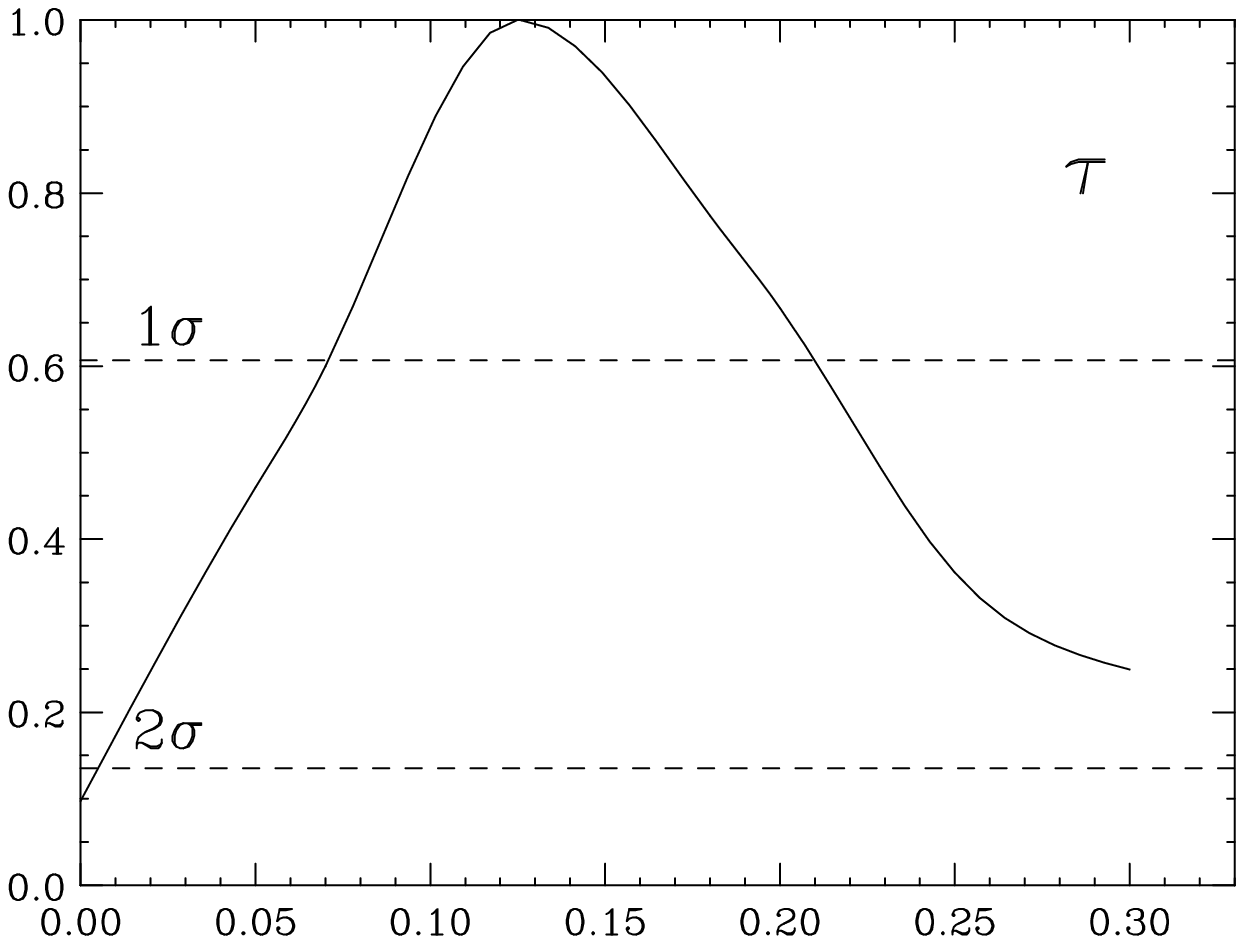,width=4cm,height=3.5cm}}
\mbox{\psfig{file=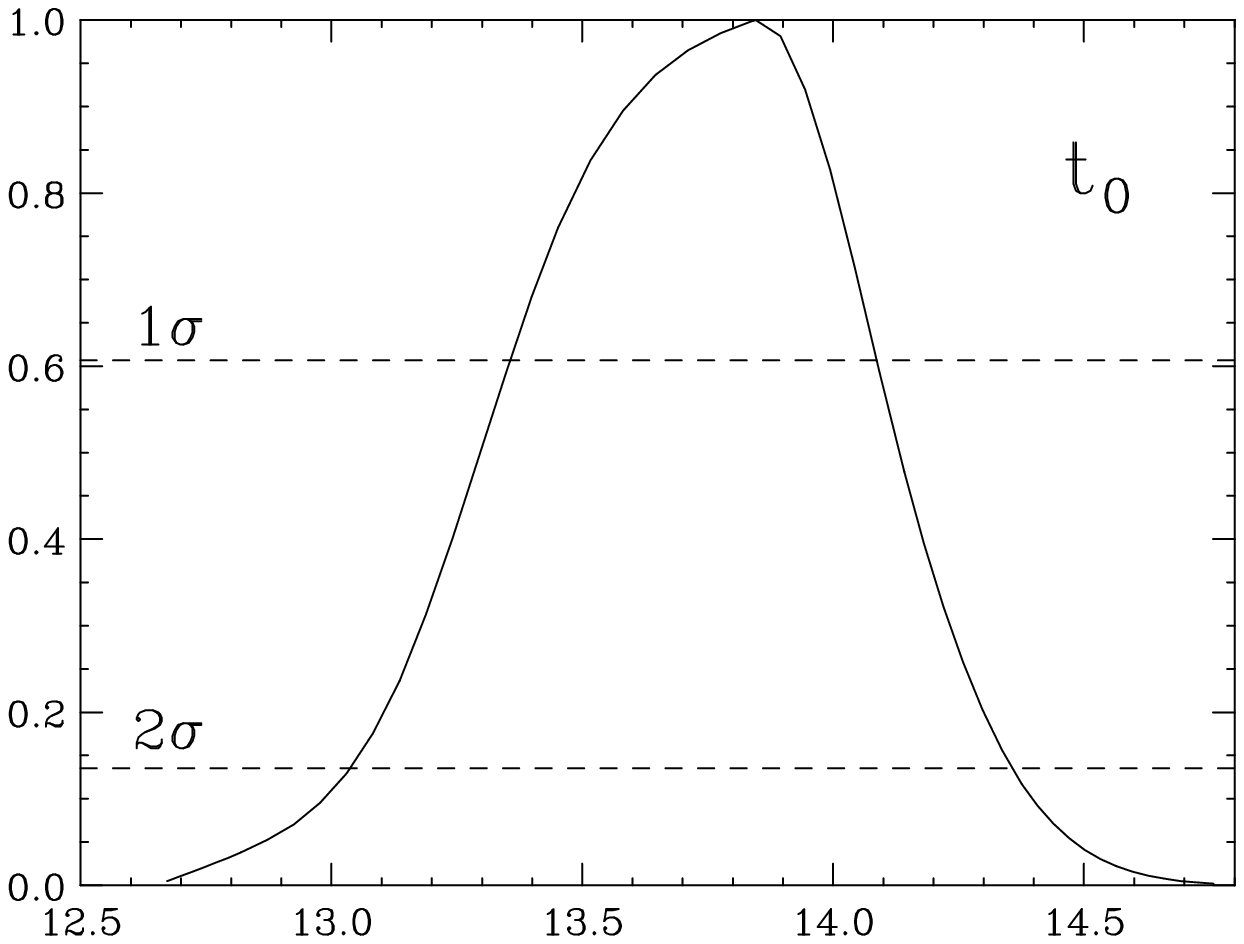,width=4.cm,height=3.5cm}
\psfig{file=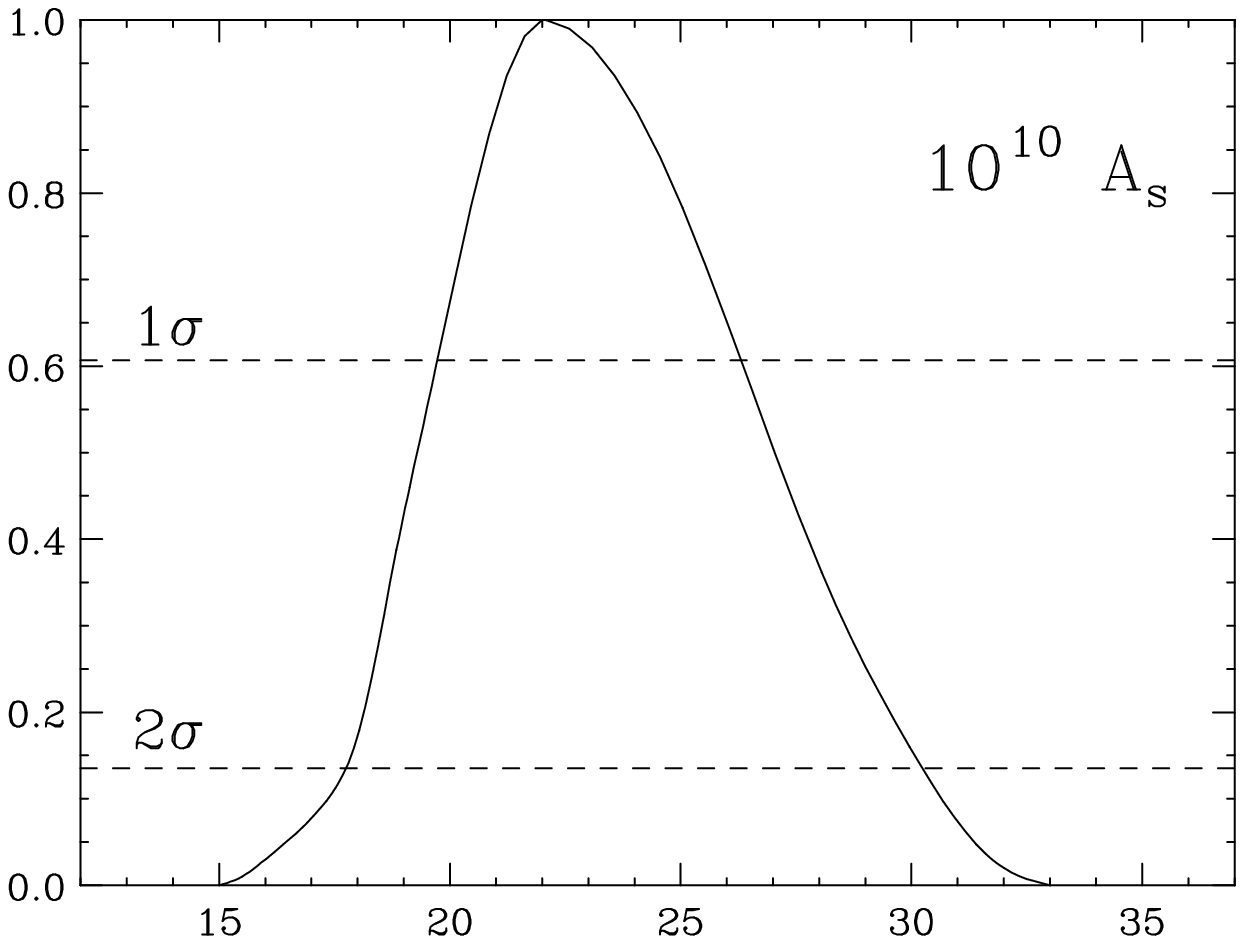,width=4cm,height=3.5cm}}
\bigskip
\caption[]{Relative likelihood plots for several cosmological parameters and 
$A_s$. $t_0$ is the age of the Universe in Gyr. We do not show the plot for
$h$ because it is not constrained by the fit;
see Table.~\ref{tab:limits}.}
    \label{fig:cosmobds}
  \end{center}
\end{figure}

The best-fit parameters are $\epsilon_1=0.016$, $\epsilon_2=-0.02$, 
$\omega_b=0.023$, $\omega_M=0.12$, $\tau=0.125$ and $h=0.78$ with 
normalization $A_s=22.0\times 10^{-10}$ with $\chi^2=1428.88$ for 1341
degrees of freedom{\footnote{To check the validity of
setting $\epsilon_i=0$, $i\geq 3$, we 
enlarged the grid to include
$\epsilon_3=\pm 0.1$ and found the minimum $\chi^2$ values to be
$\chi^2(\epsilon_3=-0.1)=1428.80$ and 
$\chi^2(\epsilon_3=+0.1)=1429.05$.
The small changes in the $\chi^2$ values
confirm that it is not necessary to include 
$\epsilon_3$ in the analysis.}}. 
The results
from Fig.~\ref{fig:cosmobds}
are in good agreement with those obtained by the WMAP
collaboration~\cite{map2}. 
This indicates that the choice of the spectral shapes does not
matter at the present precision of the WMAP data. 

\begin{figure}[ht]
  \begin{center}
\mbox{\psfig{file=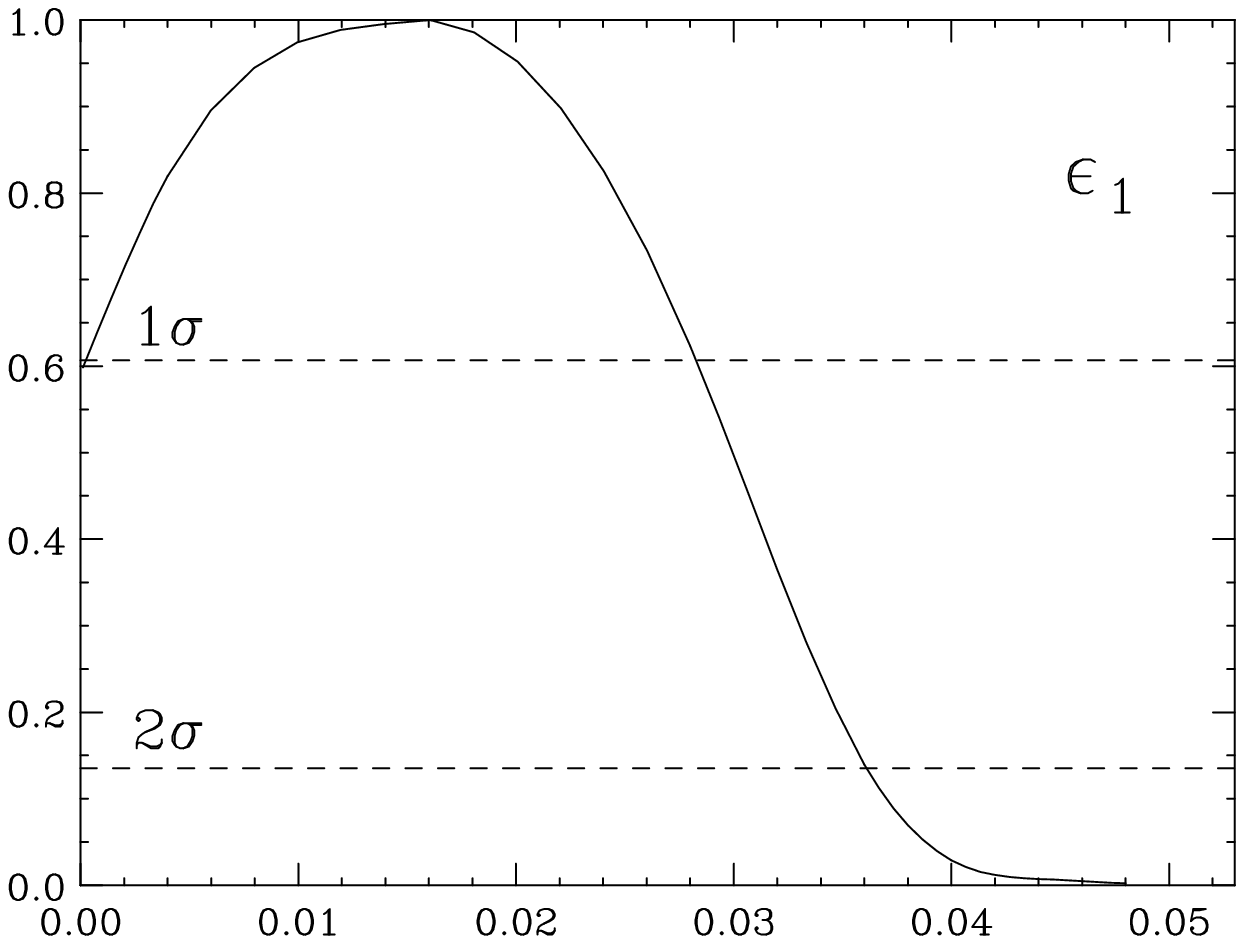,width=4cm,height=3.5cm}
\psfig{file=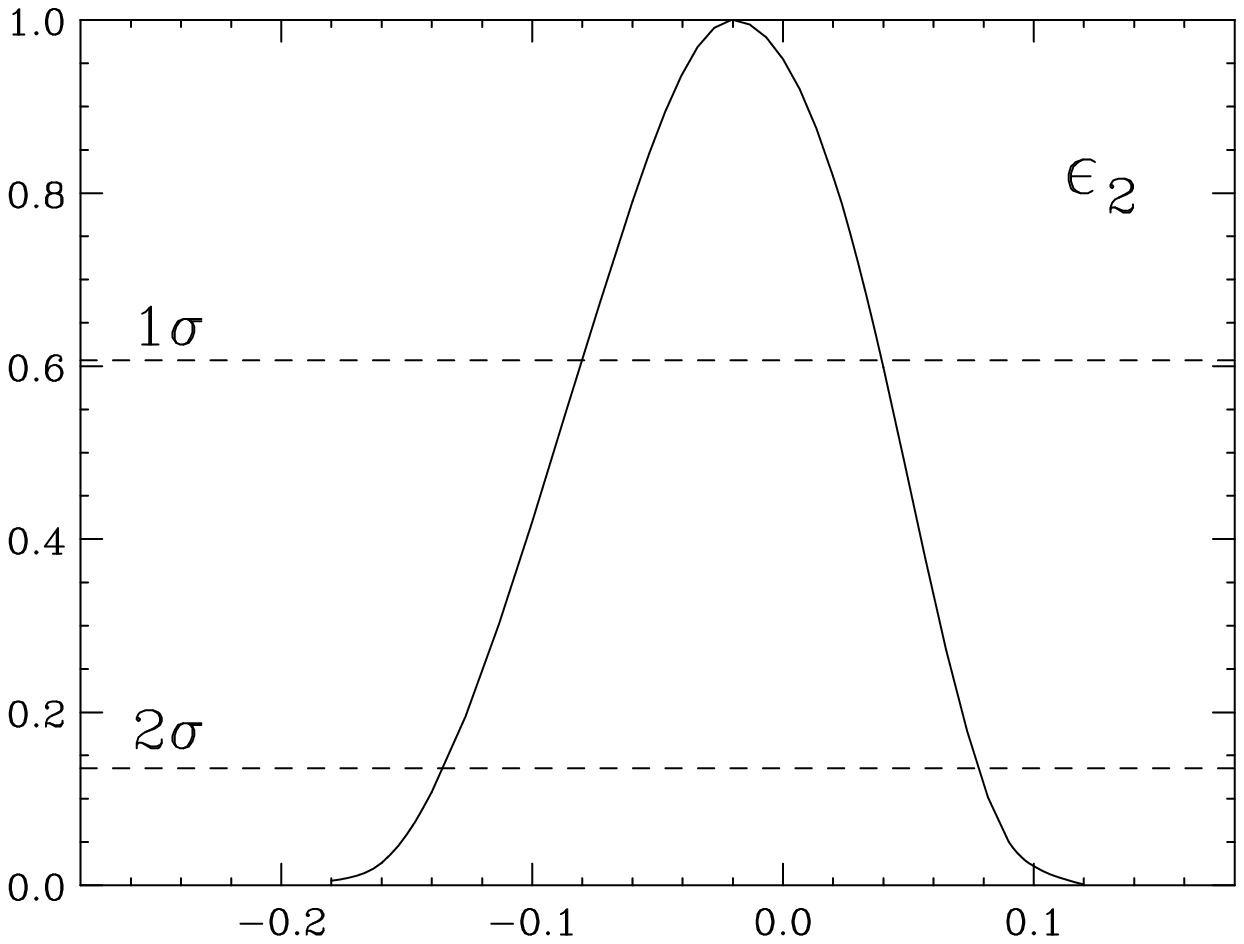,width=4cm,height=3.5cm}}
\mbox{\psfig{file=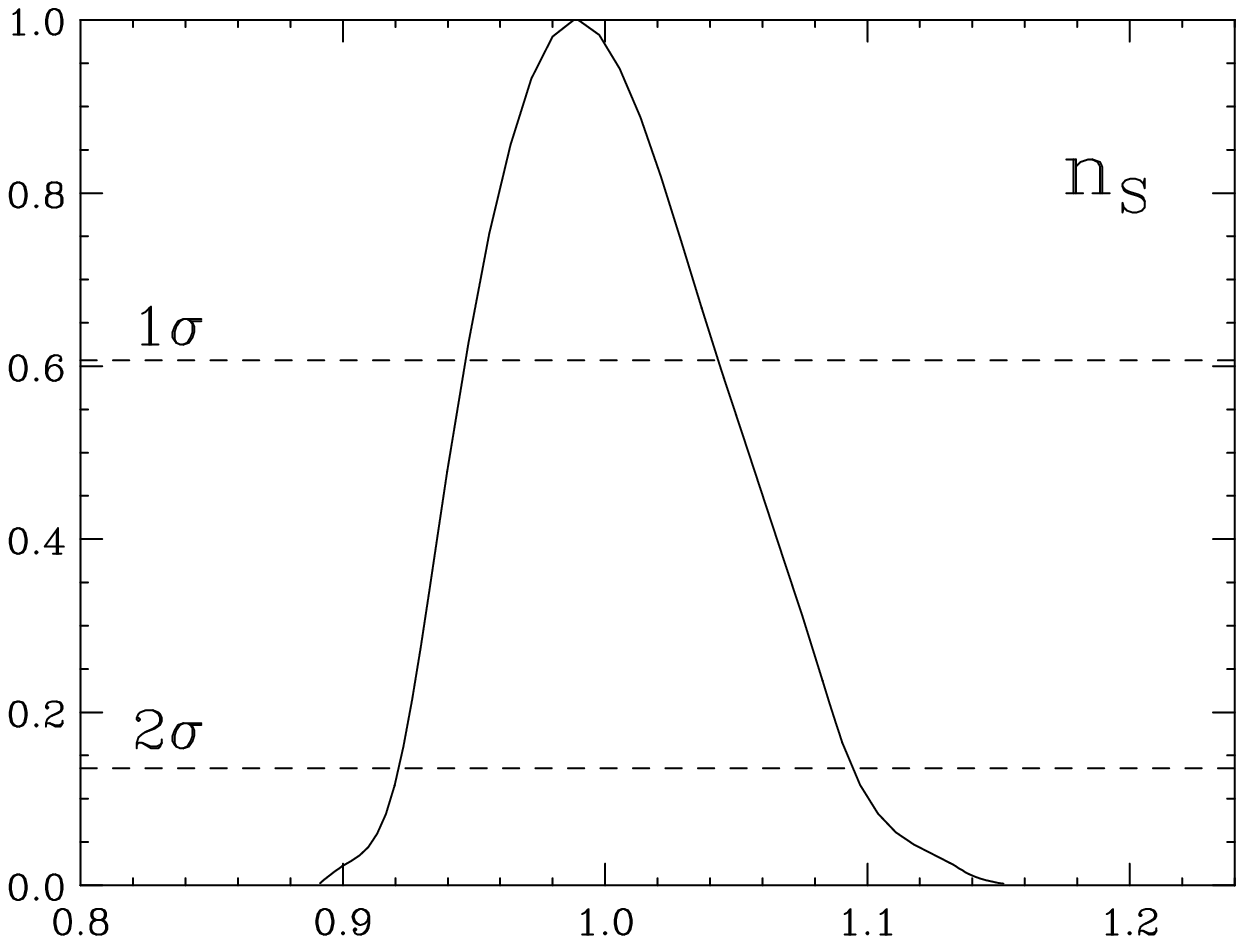,width=4cm,height=3.5cm}
\psfig{file=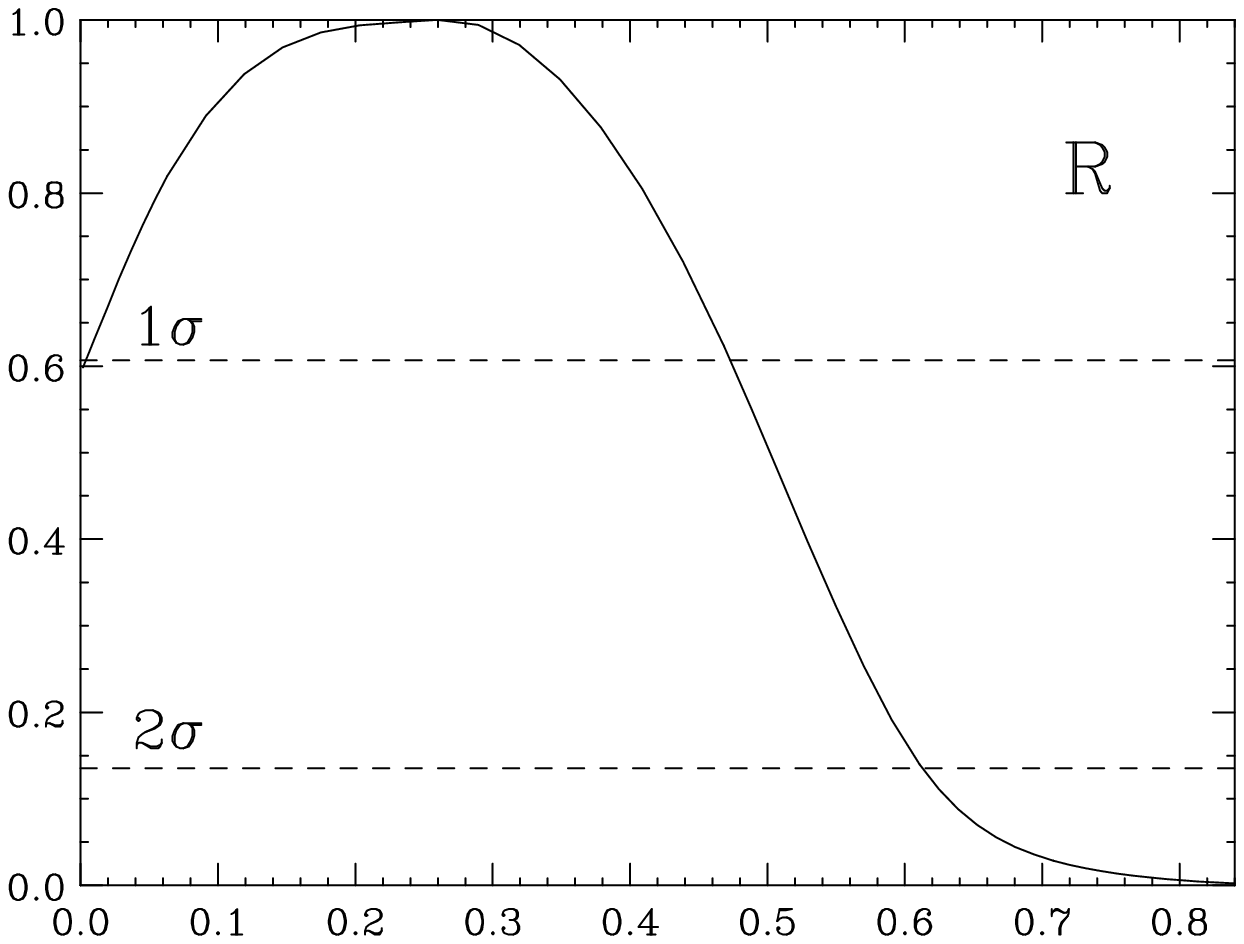,width=4cm,height=3.5cm}}
\bigskip
\caption[]{Relative likelihood plots for some inflationary parameters.}
    \label{fig:inflbds}
  \end{center}
\end{figure}

Likelihood plots for the inflationary parameters are shown in 
Fig.~\ref{fig:inflbds}. We do not show the result for $n_t$ because we 
have imposed the consistency condition, Eq.~(\ref{consistency}). 
The $1\sigma$ 
confidence limits on various parameters are provided in 
Table.~\ref{tab:limits}.
We see that the spectra are consistent with scale-invariance and with
a small tensor contribution; the best-fit 
scale-invariant spectrum with no tensor contribution has $\chi^2=1430.61$. 
Also, running of the spectral indices is insignificant.
(In our framework $\alpha_{s,t}$ are required to be consistent with  
inflationary predictions as dictated by Eqs.~(\ref{as}) and~(\ref{at}); they
are not free parameters).

\begin{table}[t]
\begin{eqnarray}
\begin{array}{c|c|c|c}
&1\sigma\ \rm{lower\ limit} &1\sigma\ \rm{upper\ limit}\\
\hline
\epsilon_1 &0 &0.028\\
\epsilon_2 &-0.08 &0.04\\
A_s\times 10^{10}&19.7  &26.3\\
\tau       & 0.07     &0.21  \\       
\omega_b   &0.022&0.026\\
\omega_M   &0.11&0.15\\
h          &0.68& -\\ \hline
n_s        &0.94 &1.04\\ 
n_t        &-0.06 &0\\
R          &0     &0.47\\
\alpha_s(\alpha_t) \times 10^3 &-0.32& 3.5\\ 
\omega_\Lambda &0.31&0.53\\
t_0/\rm{Gyr}&13.3&14.1\nonumber
\end{array}
\end{eqnarray}
\caption[]{\label{tab:limits}
The 
$1\sigma$ limits on the inflationary and cosmological parameters. Quantities 
below the line are not directly constrained by the data but derived from those
above the line. }
\end{table}

WMAP has provided important information about $R$ and the
energy scale of inflation.
\begin{equation}
  \label{eq:Rconstraint}
  R\leq 0.61\, \ \ \ \ (2\sigma\ \rm{limit})\,,
\end{equation}
and 
\begin{equation}
  \label{eq:inf1}
 {H_I\over M_{Pl}}= \sqrt{\pi \epsilon_1 A_s} \leq 1.48\times 10^{-5} \ \ \ \
(2\sigma\ \rm{limit})\,,
\end{equation}
or equivalently,
\begin{equation}
  \label{eq:inf2}
V_I^{1\over 4}=\bigg({3\over 8 \pi} {H_I^2\over M_{Pl}^2}\bigg)^{1 \over 4} M_{Pl} 
\leq 2.8\times 10^{16}\ {\rm{GeV}}\ \ \ \
(2\sigma\ \rm{limit})\,.
\end{equation}
Since $V_I^{1\over 4}\gsim 10^{15}$ GeV is consistent with 
data, it is still possible to detect inflationary gravity waves by measuring
the curl component of CMB polarization~\cite{curl}.

\vskip 0.1in
\noindent
{\bf Implications for models}:

The allowed regions of $\epsilon_1$ and $\epsilon_2$ (and
equivalently $n_s$ and $R$) are shown 
in Fig.~\ref{fig:1}. The different classes of inflationary models populate
distinct regions of the $\epsilon_2-\epsilon_1$ plane, as discussed above.
The consistency of the models with the data can be judged from the figure.
\begin{figure}[t]
  \begin{center}
\psfig{file=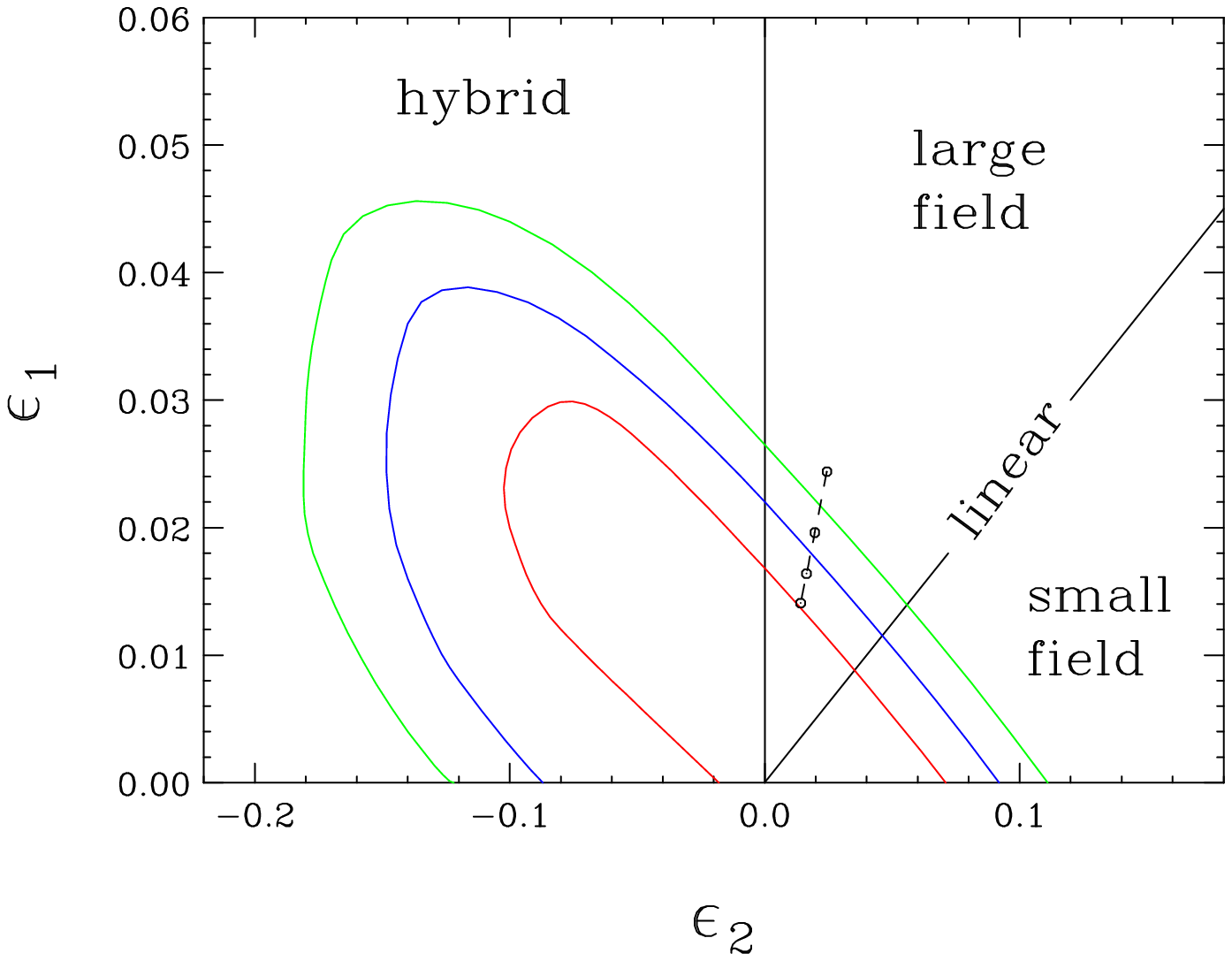,width=8cm,height=6cm}
\psfig{file=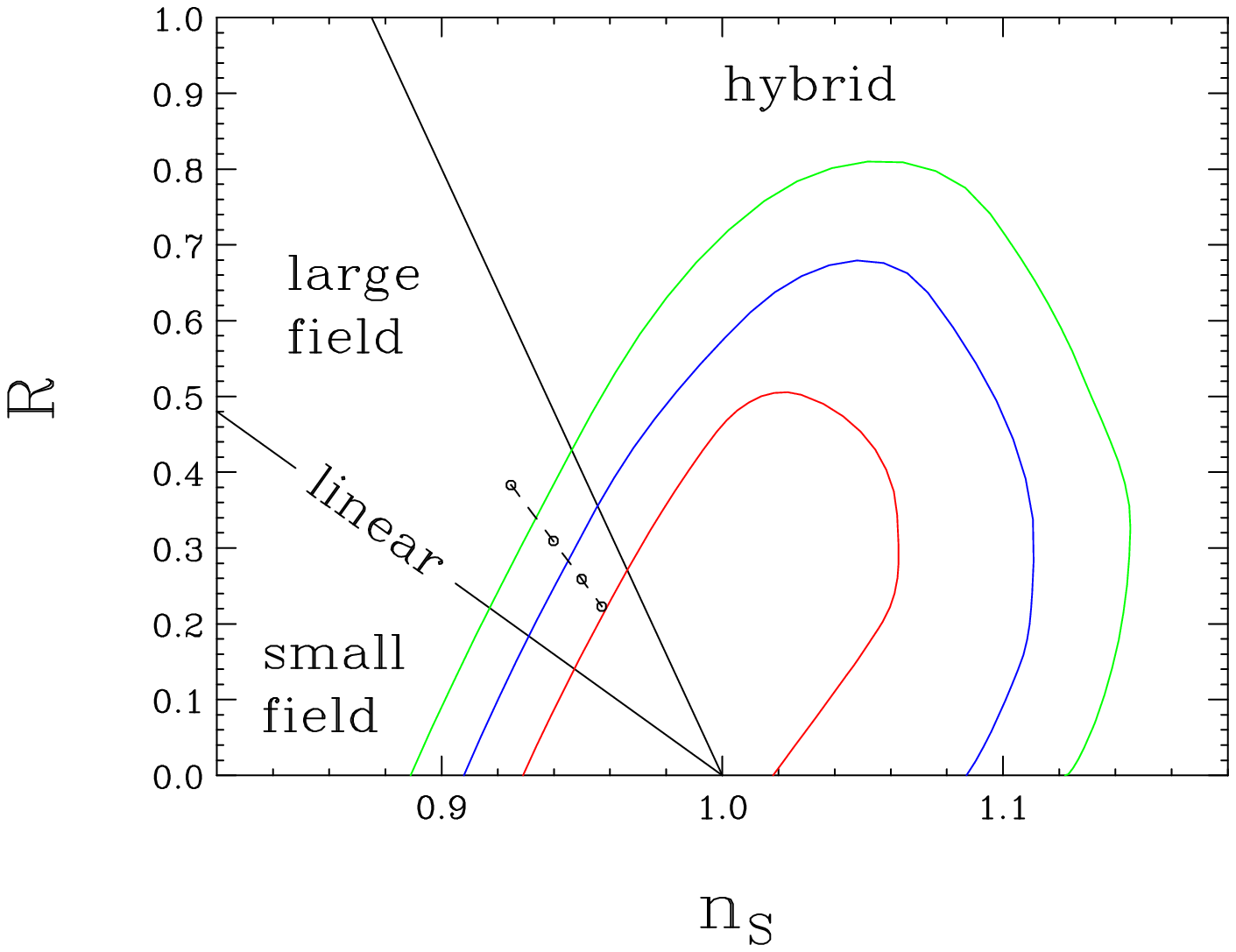,width=8cm,height=6cm}
\bigskip    
\caption[]{1$\sigma$, 2$\sigma$ and 3$\sigma$ allowed regions in the 
$\epsilon_2-\epsilon_1$ and $R-n_s$ planes. We have plotted the predictions 
for the $\lambda \phi^4$ 
potential with the number of e-folds $N=40$, 50, 60 and 70. 
The prediction approaches $\epsilon_1=\epsilon_2=0$ ($n_s=1$, $R=0$)
as $N\rightarrow \infty$.    }
    \label{fig:1}
  \end{center}   
\end{figure}
Even with the high quality of
the WMAP data, no class of models is excluded. As long as 
$\epsilon_1 \ll 1$, $\epsilon_2=0$ is consistent with data, this will remain the case.
Moreover, the allowed range for $\alpha_s$ (see Table~\ref{tab:limits}) 
 is consistent with the predictions of a wide spectrum of 
inflationary models, and so
does not help in discriminating between them. 

We now place some constraints on large-field and small-field 
models whose predictions do not involve
too much freedom.

For the monomial potentials ($p\geq 2$) of large-field models,
\begin{eqnarray}
  \label{eq:mon}
\epsilon_1&\simeq&{p \over 4}\, \epsilon_2\,,\\
  N&\simeq&{p \over 4}\bigg({1\over \epsilon_1} -1\bigg)\,,
\end{eqnarray}
where $N$ is the number of 
e-folds of inflation from the time that scales probed by the CMB leave
the horizon until the end of inflation{\footnote{We are reverting to the
conventional definition in which the number of e-folds is counted 
backward in time; in the definition of the horizon-flow parameters, e-folds
are counted forward in time.}. At least about 40 e-folds are needed
for the Universe to be flat and homogeneous, and typically the largest 
value is 70~\cite{n}. Since 
$\epsilon_1 \ll 1$, $\epsilon_2=0$ is allowed, $p$ cannot be constrained
independently of $N$. 

The $3\sigma$ exclusion of the $\lambda \phi^4$ 
potential in Ref.~\cite{map1} was based on an analysis of WMAP data 
in combination with higher $l$ CMB data and large scale structure data,
assuming $N=50$ 
(for which $n_s=0.94$ and $R=0.32$). Their results are 
shown in the second row of their Fig.~4. 
Note that the point $n_s=0.94$, 
$R=0.32$ lies inside
the $3\sigma$ region of our Fig.~\ref{fig:1} and is therefore not excluded by
 WMAP data alone. If instead $N$
is 60, then $n_s=0.95$ and $R=0.27$; this point lies in the 95\% C.~L. 
allowed region of the second row of Fig.~4 of Ref.~\cite{map1} and 
within the $2\sigma$ region of Fig.~\ref{fig:1}.

If $\epsilon_1\neq 0$ ($\epsilon_2> 0$), 
a lower bound (upper bound) 
$p/4>\epsilon_1^{min}/\epsilon_2^{max}$ 
($p/4<\epsilon_1^{max}/\epsilon_2^{min}$) ensues; $p$ is determined if both
conditions on the $\epsilon$'s are simultaneously satisfied.

Since we expect $p\ll 4N$, 
\begin{equation}
  \label{eq:Nbound}
  N\gsim {1 \over \epsilon_2^{max}}=12.5\, \ \ \ \ (2\sigma\ \rm{limit})\,,
\end{equation}
and 
\begin{equation}
  \label{eq:pbound}
p\lsim 4N\epsilon_1^{max}=0.15N\, \ \ \ \ (2\sigma\ \rm{limit})\,.
\end{equation}
 
For small-field models with $p\geq 3$~\cite{review}, 
\begin{equation}
  \label{eq:1}
  \epsilon_1\ll \epsilon_2={2\over N} {p-1\over p-2}
\end{equation}
The least stringent bound on $N$ occurs in the limit 
$p\rightarrow \infty$,
\begin{equation}
  \label{eq:2}
  N\geq{2 \over \epsilon_2^{max}}=25\, \ \ \ \ (2\sigma\ \rm{limit})\,.
\end{equation}

For the small-field quadratic potential ($p=2$),
\begin{equation}
  \label{eq:3}
  \epsilon_1\ll \epsilon_2={1\over 2 \pi} {M^2_{Pl}\over \mu^2}\,.
\end{equation}
We find the $2\sigma$ bound
\begin{equation}
\mu > M_{Pl}/\sqrt{2\pi \epsilon_2^{max}}=1.4 M_{Pl}\,.
\end{equation}

Similar constraints can be placed on other models, but unfortunately, 
they are not very enlightening.

\vskip 0.1in
\noindent
{\bf Conclusions}:

WMAP has provided compelling  evidence for the inflationary paradigm.
We have adopted the explicit predictions
of single-field slow-roll inflation for the shapes of the power-spectra to
analyze WMAP data. 
The fact that 
our parameter determinations are consistent with those obtained with the
standard power-law parameterization by the
WMAP collaboration provides further evidence for slow-roll inflation. 
Since exact scale-invariance and a negligible
tensor contribution to the density perturbations are adequate to describe
the data, it is not presently possible to exclude classes of inflationary
models. 

We have shown how different classes of inflationary models
can be distinguished in the $\epsilon_2-\epsilon_1$ plane of the horizon-flow
parameters.
If and when the horizon-flow parameters 
$\epsilon_1$ or/and $\epsilon_2$ are determined to be
non-zero, large numbers of inflationary models will be ruled out. 
For that, we await even higher precision data from WMAP
and eventually from Planck. 

\vskip 0.1in
{\it Acknowledgments}:
The computations were carried out on the CONDOR system at the University
of Wisconsin, Madison with parallel processing on up to 200 CPUs. We thank
S. Dasu, W. Smith, D. Bradley and S. Rader for providing access and assistance
with CONDOR. We thank L. Verde for communications regarding the WMAP 
likelihood code. 
This research was supported by the U.S.~DOE  
under Grants No.~DE-FG02-95ER40896 and No.~DE-FG02-91ER40676, by the NSF
under Grant No.~PHY99-07949
and by the Wisconsin Alumni Research Foundation. VB and DM thank the Kavli
Institute for Theoretical Physics at the University of California, Santa
Barbara for hospitality.

\end{document}